\documentclass[aps,prl,reprint,superscriptaddress,nofootinbib,longbibliography]{revtex4-2}

\usepackage{ORCIDinREVTeX}

\usepackage[T1]{fontenc}
\usepackage{xcolor, soul}
\definecolor{darkgreen}{rgb}{0.0,0.55,0.0}
\usepackage{graphics}
\usepackage{amssymb,bm}
\usepackage{minibox}
\usepackage[colorlinks=true, allcolors=darkgreen]{hyperref}
\usepackage{empheq}
\usepackage{enumerate}
\usepackage{multirow}
\usepackage{ulem}
\usepackage{float}
\usepackage{cancel}
\usepackage{soul}

\usepackage{cleveref}
\usepackage{relsize}

\usepackage[ddmmyyyy]{datetime}

\usepackage[tmargin=1.5cm, bmargin=1.5cm, lmargin=1.5cm, rmargin=1.5cm]{geometry}

\usepackage{lmodern}
\usepackage{enumitem}

\newcommand{\nc}{\newcommand}

\nc{\non}{\nonumber}
\nc{\hsp}{\hspace{0.5cm}}
\nc{\lsp}{\hspace{1cm}}
\nc{\Lsp}{\hspace{2cm}}
\nc{\LLsp}{\lsp\lsp}
\nc{\lra}{\longrightarrow}
\nc{\p}{\prime}
\nc{\dd}{\mathrm{d}}
\nc{\sgn}{\text{sgn}}
\nc{\ph}{\varphi}
\nc{\op}{{\cal O}}
\nc{\cL}{{\cal L}}
\nc{\tr}{{\text{Tr}}}
\nc{\eq}{\text{Eq.~}}
\nc{\cg}{{\cal G}}
\nc{\ch}{{\bm h}}
\nc{\cZ}{\mathbb Z}
\nc{\cw}{\cos\theta_{\textsc w}}
\nc{\sw}{\sin\theta_{\textsc w}}
\nc{\cwsq}{\cos^2\theta_{\textsc w}}
\nc{\swsq}{\sin^2\theta_{\textsc w}}
\def\zBB{{\mathbbm Z}}
\def\z2{\zBB_2}
\nc{\beq}{\begin{equation}}  \nc{\eeq}{\end{equation}}
\nc{\bea}{\begin{eqnarray}}  \nc{\eea}{\end{eqnarray}}
\nc{\baa}{\begin{array}}     \nc{\eaa}{\end{array}}
\nc{\bit}{\begin{itemize}}   \nc{\eit}{\end{itemize}}
\nc{\ben}{\begin{enumerate}} \nc{\een}{\end{enumerate}}
\nc{\bce}{\begin{center}}    \nc{\ece}{\end{center}}
\nc{\bpm}{\begin{pmatrix}}   \nc{\epm}{\end{pmatrix}}
\nc{\bvt}{\begin{verbatim}}  \nc{\evt}{\end{verbatim}}
\def\lsim{\mathrel{\raise.3ex\hbox{$<$\kern-.75em\lower1ex\hbox{$\sim$}}}}
\def\gsim{\mathrel{\raise.3ex\hbox{$>$\kern-.75em\lower1ex\hbox{$\sim$}}}}

\def\udots{\mathinner{\mkern1mu\raise1pt\vbox{\kern7pt\hbox{.}}\mkern2mu\raise4pt\hbox{.}\mkern2mu\raise7pt\hbox{.}\mkern1mu}}
\def\<#1>{\mathinner{\langle#1\rangle}}

\nc{\hc}{\hbox {H.c.}}
\nc{\noi}{\noindent}
\nc{\barx}{\bar{x}}
\nc{\pbarn}{\;\hbox {pb}}
\nc{\fbarn}{\;\hbox {fb}}

\def\gev{\;\hbox{GeV}}

\def\mpl{M_{\rm Pl}}

\definecolor{agray}{rgb}{0.95, 0.95, 0.99}

\def\sm{\textsc{\rm SM}}
\def\dm{\textsc{\rm DM}}

\def\cxh{{\cal C}_{X}^{\bm h}}
\def\cxp{{\cal C}_{X}^\phi}
\def\ghphi{g_{h\phi}^{}}
\def\mx{m_X}

\def\eq#1{Eq.~(\ref{#1})}
\def\fig#1{Fig.~\ref{#1}}

\def\rcite#1{Ref.~\cite{#1}}
\definecolor{MyBlue}{RGB}{61,80,210}
\definecolor{MyRed}{RGB}{172,18,10}
\definecolor{MyGreen}{RGB}{56,92,41}

\begin{document}

\title{Implications of time-dependent inflaton decay on reheating and dark matter production}

\author{Aqeel Ahmed}
\orcid{0000-0002-2907-2433}
\email{aqeel.ahmed@mpi-hd.mpg.de}
\affiliation{Max-Planck-Institut für Kernphysik (MPIK), Saupfercheckweg 1, 69117 Heidelberg, Germany}

\author{Bohdan Grzadkowski}
\orcid{0000-0001-9980-6335}
\email{bohdan.grzadkowski@fuw.edu.pl}

\author{Anna Socha}
\orcid{0000-0002-4924-9267}
\email{anna.socha@fuw.edu.pl}
\affiliation{Faculty of Physics, University of Warsaw, Pasteura 5, 02-093 Warsaw, Poland}

\date{\today}

\begin{abstract}
We discuss the production of radiation and dark matter assuming a time-dependent inflaton decay rate during the reheating period.
It is shown that the time dependence of the inflaton decay rate can substantially modify the reheating dynamics.
As an illustration, a leading interaction between the inflaton $\phi$ and the Higgs doublet ${\bm h}$ of the form $\phi |{\bm h}|^2$ was adopted.
In the presence of such interaction, the Higgs doublet acquires a $\phi$-dependent mass which generates vacuum-expectation-value due to inflaton oscillations and breaks the Standard Model gauge symmetry. 
This leads to a time-dependent inflaton decay rate during the reheating period, and consequently, the production of radiation and dark matter during this period is modified. 
Regions of the parameter space that describe the observed value of the dark matter abundance were found and compared with the standard case when inflaton the decay rate is constant.
\end{abstract}
\maketitle

\paragraph{Introduction.}
\label{Introduction}
One of the most outstanding mysteries of high energy physics is the lack of understanding of the nature and origin of dark matter (DM), which constitutes nearly 85\% of the observed matter density in our Universe~\cite{Aghanim:2018eyx}. 
Usually, it is assumed that DM couples to the Standard Model (SM) with the corresponding coupling being of the order of the weak strength, for a review see~\cite{Arcadi:2017kky}; however, it should be emphasized that the only interaction between 
these two sectors confirmed experimentally is gravity.
In recent years considerable interest has been observed in building DM models where the DM interacts with the SM through gravity only.         
Such models offer several rather natural mechanisms which can explain production of DM.
For example, DM production during inflation period in the early Universe through quantum fluctuations~\cite{Chung:1998zb,Graham:2015rva,Ema:2018ucl,Alonso-Alvarez:2018tus,Hashiba:2018tbu,Li:2019ves,Ema:2019yrd,Tenkanen:2019aij,Alonso-Alvarez:2019ixv,
Ahmed:2020fhc,Ahmed:2019mjo,Kolb:2020fwh,Ling:2021zlj,Arvanitaki:2021qlj}, the graviton mediated production
of DM from the annihilation of the SM particles ~\cite{Garny:2015sjg,Tang:2017hvq,Garny:2017kha,Mambrini:2021zpp} or in the background of the inflaton field  ~\cite{Mambrini:2021zpp,Haque:2021mab,Clery:2021bwz,Haque:2022kez,Clery:2022wib}, and production of DM through Planck scale 
suppressed portals to SM~\cite{Kolb:2017jvz,Bernal:2018qlk,Chianese:2020yjo,Chianese:2020khl}.
The observed DM relic abundance can be achieved within these models for a wide range of DM masses from sub eV to the Planck mass. 
One aspect common in different production mechanisms of gravitational DM is that such production should occur in the very early Universe at the late stages of inflation. 
Usually, the relevance of the inflationary and reheating dynamics is ignored with a minimal assumption, such as instantaneous reheating.  However, it is well known that the inflaton potential sets the initial conditions for the reheating phase which could last longer with a generic equation of state $w$~\cite{Shtanov:1994ce,Garcia:2020wiy,Redi:2020ffc} affecting DM production~\cite{Giudice:2000ex,Chung:1998rq,Harigaya:2014waa,Chen:2017kvz,Bernal:2019mhf,Garcia:2020eof,Mambrini:2021zpp}. 
Even in the case of non-instantaneous reheating, it is assumed that the inflaton perturbative decay rate $\Gamma_\phi$ is constant, and when the Hubble scale $H$ becomes of the order of $\Gamma_\phi$ the reheating phase ends. 

However, this widely used assumption of constant $\Gamma_\phi$ can be violated in generic models of perturbative reheating, e.g., when the inflaton interacts with the SM Higgs field or inflaton has non-trivial potential, the latter was discussed recently in \rcite{Garcia:2020wiy}, see also~\cite{Barman:2022tzk}.
In this Letter, we first consider a general case where the inflaton decay rate $\Gamma_\phi$ is a time-dependent function and study its non-trivial consequences for the SM energy density during the reheating phase. 
Then we present a model where such time-dependent inflaton decay rate is achieved with an interaction of the form $\phi |{\bm h}|^2$ between the inflaton $\phi$ and the Higgs doublet ${\bm h}$. 
In this case, the time dependence of the inflaton decay rate appears through the production of Higgs-boson pairs with time/$\phi$-dependent mass.
We note that such an effect appears naturally in all models of reheating, which is sourced by a spin-$0$ field; therefore, it is common to nearly all scalar field inflationary scenarios.
In the last part of this Letter, adopting the paradigm of DM interacting through Planck suppressed operators, we consider the implications of the proposed non-trivial reheating for a freeze-in production of DM. 
For concreteness, we consider a model of vector DM, $X_\mu$, which is a gauge boson of an Abelian $U(1)_{X}$ symmetry.
With Planck scale suppressed DM interactions, the vector DM is either produced by the inflaton/Higgs decay or through the freeze-in mechanism via graviton or Higgs boson exchange. 

\paragraph{Reheating with a time-dependent inflaton decay.}
\label{Reheating with a time-dependent inflaton decay}
The reheating dynamics is captured by the following Boltzmann equation for the inflaton, 
\begin{align}
\dot{\rho}_\phi + 3(1+w)  H \rho_\phi &= - \Gamma_\phi(a) \rho_\phi  \,,\label{eq:beqn_phi} 
\end{align}
where $\rho_\phi$ and $\Gamma_\phi (a)$ should be interpreted as time averaged~\cite{Shtanov:1994ce} inflaton energy density and inflaton ``decay width''~\cite{Ichikawa:2008iq} \,\footnote{The inflaton ``decay width'' describes a process of quantum production of radiation from a vacuum in the presence of the classical inflaton field, see~\cite{Ichikawa:2008iq}.}, respectively. 
The {\it overdot} denotes derivative w.r.t. the cosmic time $t$, $w \equiv p_\phi/\rho_\phi$ is assumed to be a constant equation-of-state parameter ($p_\phi$ is the corresponding pressure), and $H \equiv \dot a/ a$ is the Hubble parameter with $a$ being the scale factor of the FLRW metric. 
We assume that $\Gamma_\phi(a)$ is much smaller than the Hubble rate $H(a)$ in the early stages of reheating.  Furthermore, during this phase, the total energy density is dominated by the inflaton energy density, i.e., \mbox{$\rho_{\rm tot}(a)\simeq\rho_\phi(a)$}, such that
\beq
\rho_\phi(a)\simeq 3\mpl^2 H^2 \approx 3\mpl^2 H_e^2 \Big(\frac{a_e}{a}\Big)^{3(1+w)}. 
\eeq
Above $\mpl=2.4\times 10^{18}\gev$ is the reduced Planck mass, and $H_e$ denotes the Hubble scale at the end of inflation marked by an arbitrary scale factor $a=a_e$. 
During the reheating phase, we assume that the inflaton coherently oscillates and that its decay rate is parametrized as~\footnote{In this work, we limit ourselves to time-dependence of inflaton decay rate with power-law of the scale factor. However, the analysis can be extended to more general time-dependent inflaton decay rates.}, 
\beq
\Gamma_\phi (a) = \Gamma_\phi^e\Big(\frac{a_e}{a}\Big)^{\beta},
\eeq
where $\Gamma_\phi^e$ is a time-independent inflaton width and $\beta$ is a constant parameter. The end of reheating phase is defined as when inflaton energy density is equal to that of the SM, i.e. $\rho_{\phi}=\rho_{\rm SM}^{}$ at $a=a_{\rm rh}$, which roughly occurs when $\Gamma_\phi(a_{\rm rh})\sim H(a_{\rm rh})\equiv H_{\rm rh}$.
During the reheating phase, the SM energy density is governed by the following Boltzmann equation,
\begin{align}
\dot{\rho}_{\rm SM}^{} + 4 H \rho_{{\rm SM}}^{} \!&=\! \Gamma_\phi (a) \, \rho_\phi.
 \label{eq:be_sm} 
\end{align}
It is straightforward to solve \eq{eq:be_sm} as, 
\begin{align}
 \rho_{{\rm SM}}^{} (a)\!&=\! \frac{6\mpl^2 H_e \Gamma_\phi^e}{5\!-\!3w\!-\!2\beta}  \bigg[\Big(\frac{a_e}{a}\Big)^{\beta+3 (1 + w)/2}-\Big(\frac{a_e}{a}\Big)^4\bigg],
 \label{eq:rho_sm} 
\end{align}
where we require $\beta\!\leq\!(5 \!-\! 3 w)/2$ such that the first term in the above square bracket is the dominant term for $a_e\!<\!a\!\leq\! a_{\rm rh}$. After the end of reheating, the SM energy density scales as in the standard radiation-dominated Universe, i.e., $\propto a^{-4}$. 
The Hubble rate at the end of reheating is, $H_{\rm rh}\!=\!2\Gamma_\phi(a_{\rm rh})/(5\!-\!3w\!-\!2\beta)$.
Note the non-trivial power-law dependence of SM energy density on $\beta$ parameter in \eq{eq:rho_sm} is one of the important results of this work. For instance, for $w\!=\!0$ and $\beta\!=\!-3/2$, we find that the SM energy density remains roughly constant during the reheating phase, which implies that the temperature during this period stays constant as well. Such non-trivial scaling of SM energy density can have significant implications for models of ultraviolet freeze-in as they are highly sensitive to the temperature evolution during the reheating phase.  In the following, we consider a specific model that realizes the proposed non-trivial feature of the time-dependent inflaton decay rate, and then we investigate its implications for the SM and DM phenomenology. 
\paragraph{Example model.}
\label{inflaton}
As an illustration, we consider the $\alpha$-attractor T-model of inflation~\cite{Kallosh:2013hoa,Kallosh:2013yoa} such that the inflaton potential is written as
\beq
\begin{aligned}
V(\phi) &\!=\!\Lambda^{4} \tanh ^{2 n}\!\bigg(\frac{|\phi|}{M}\bigg) \!\simeq\! \begin{cases}
\Lambda^{4}\,,  & |\phi| \gg\! M\\
\Lambda^{4}\left|\frac{\phi}{M}\right|^{2 n}\,,  & |\phi| \ll\! M
\end{cases},
\end{aligned} \label{eq:inf_pot}
\eeq
where the parameter $\Lambda$ determines the scale of inflation, whereas $M$ is related to the Planck mass $\mpl$ with the $\alpha$ parameter of the $\alpha$-attractor T-model as $M\equiv\sqrt{6 \alpha}\, \mpl$.
For the inflaton field values $\phi\gtrsim M$, the inflaton potential is nearly constant, whereas the end of inflation occurs for $\phi\sim M$, such that during the reheating phase, the inflaton potential has the polynomial form, i.e., $\phi^{2n}$.
In what follows, we will consider only the case $n\!=\!1$, which implies $w\!=\!0$ during reheating epoch (for general case see~\cite{Ahmed:2022tfm} and also see~\cite{Ahmed:2022qeh}). 

The classical equation of motion (EoM) for the inflaton field in the FLRW metric is given by
\beq
\ddot{\phi}+3 H \dot{\phi}+ V^\prime(\phi)=0	\label{eq:inf_eom}\,,
\eeq
where {\it prime} denotes derivative of the potential w.r.t. $\phi$.
Note that the above equation implies the time averaged Boltzmann equation (\ref{eq:beqn_phi}) for $\Gamma_\phi\ll H$.
Furthermore, \eq{eq:inf_eom} neglects possible interaction with the SM and DM sectors.
During the reheating phase, the general solution to \eq{eq:inf_eom} can be recast in terms of a slowly-varying envelope $\varphi(t)$ and a fast-oscillating function ${\cal P}(t)$ as \cite{Shtanov:1994ce,Garcia:2020wiy},
\begin{align}
\phi(t) = \varphi(t) \cdot \mathcal{P}(t), \label{eq:phi_app}
\end{align}
where $\varphi$ is defined through $\rho_{\phi} \equiv V(\varphi)=\Lambda^4|\varphi/M|^{2n}$.
For $n=1$, the envelope $\varphi$ as a function of scale factor $a$ reads
\begin{align}
\varphi(a) &= \varphi_e\, \Big(\frac{a_e}{a} \Big)^{3/2}  \;\; \text{with} \;\; &\varphi_e&\equiv \frac{M}{2}\,{\rm sinh^{-1}}\!\Big(\! \frac{2}{\sqrt{3\alpha}}\Big), \label{eq:envelope_sol}
\end{align}
where $\varphi_e$ is a value of the inflaton field at the end of inflation, defined by the condition imposed on the slow-roll parameter $\epsilon_V (a_e) \!\equiv\! 1/2 \, (\mpl V^\prime(\phi)/V(\phi))^2 \!=\!1$. 
Whereas, for the quadratic inflaton potential, the fast-oscillating function takes a simple form~\cite{Kofman:1997yn},
\begin{align}
\mathcal{P}(t) &= \cos( m_\phi\, t )\,,	\quad	&m_{\phi}^{2}& \equiv\frac{\partial^2 V(\phi)}{\partial \phi^2}\bigg\vert_{\phi=\varphi}
\!\!\!=  \frac{2\Lambda^4}{M^2},	\label{eq:mphi}
\end{align}
where $m_\phi$ denotes the effective inflaton mass, such that the period of fast-oscillation is $2\pi/m_\phi$.

\begin{figure}[t!]
\begin{center}
\includegraphics[width=\linewidth]{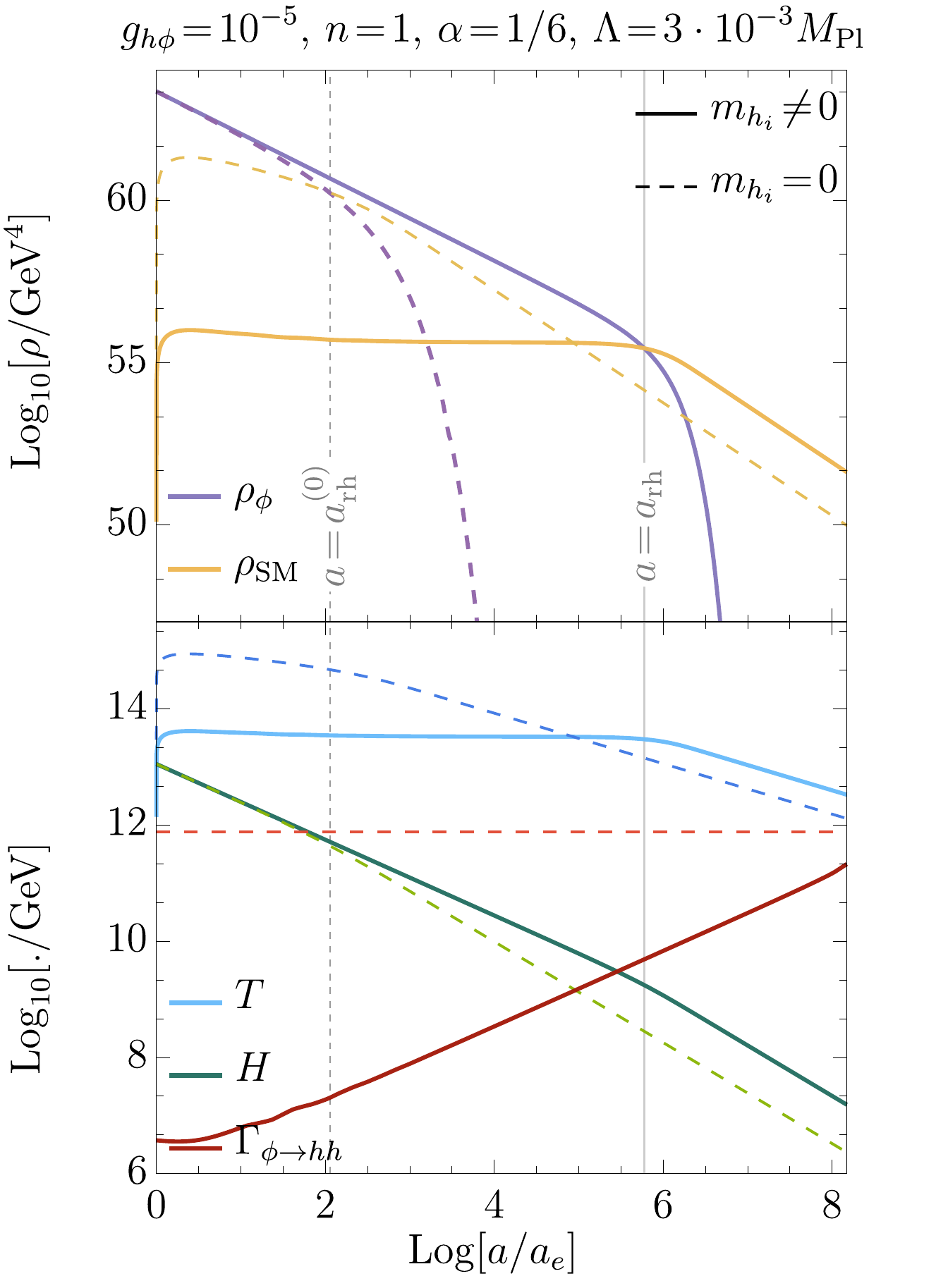}\vspace{-13pt}
\caption{Inflaton and radiation energy densities are plotted in the upper panel as a function of the scale factor $a$. The lower panel presents the evolution of the thermal bath temperature $T$, the Hubble parameter $H$, and the averaged inflaton width $\Gamma_{\phi \rightarrow hh} $ for the massless (dashed curves) and massive (solid curves) cases. The vertical solid (dashed) gray line indicates the end of reheating, i.e., $\rho_\phi=\rho_{\rm SM}^{}$ when the produced Higgs boson pairs are massive (massless), i.e., inflaton decay rate is time-(in)dependent. 
}
\label{fig:plotn1gh5}
\end{center}
\end{figure}

During the reheating phase, the inflaton coherently oscillates and transfers its energy density to the SM sector. The lowest dimensional interaction between the inflaton and SM is through the Higgs portal,
\beq
\cL\supset -g_{h\phi}^{} \mpl\, \phi |{\bm h}|^2,	 \label{eq:Lhphi}
\eeq
where we assume the dimensionless coupling $g_{h\phi}^{}\!>\!0$. 
We neglect higher dimensional inflaton interactions with the SM fermions and gauge bosons, favoring the above dominant interaction. 
Perturbativity sets an upper-bound on $\ghphi$ while requiring Higgs stability during inflation implies a lower-bound on $\ghphi$~\cite{Ahmed:2022tfm}, such that
\begin{align}
\frac{\Lambda^2}{\phi\mpl} \gtrsim \ghphi&\gtrsim\frac{3}{4}\sqrt{6 \alpha}\bigg( \frac{\Lambda^2}{\phi \mpl} \bigg)^2 \frac{\phi}{M}. \label{ghphi_limit}
\end{align}  
Above, we assume $\Lambda$ to be the UV-scale that governs the reheating dynamics.
During this phase, the Higgs effective potential is given by 
\begin{align}
V({\bm h}) &=
\mu_h^2(\phi) \,|{\bm h}|^2+\lambda_h\, |{\bm h}|^4\,,
\label{eq:H_pot}
\end{align}
where the Higgs mass parameter is a function of rapidly oscillating inflaton field through \eq{eq:Lhphi}, i.e.
\beq
\mu_h^2(\phi) \equiv g_{h\phi}^{} \,\mpl\, \phi(t)=g_{h\phi}^{}\, \mpl\, \varphi(t)\,\mathcal{P}(t),	\label{eq:muh}
\eeq
where we have neglected the constant contribution to the Higgs mass proportional to the electroweak vev $v_{\tiny{\rm  EW}} \simeq 246\; {\rm GeV}$. 
Furthermore, we note that thermal corrections to the SM Higgs mass term \mbox{$\Delta \mu_h^2\sim 0.4\, T^2\sim 0.4 \sqrt{\rho_{\rm SM}(a)}\ll \mu_h^2(\phi)$} for the parameter space where inflaton decay rate is time-dependent, for details see~\cite{Ahmed:2022tfm}. 
Due to the inflaton oscillations during reheating, the Higgs field goes through rapid phase transitions with $\phi$-dependent vev:
\begin{align}
v_h^2(\phi) &=\begin{dcases}
~0\,, & \mathcal{P}(t)> \! 0\,,		\\
\!|\mu_h^2(\phi)|/\lambda_h\,,	\quad& \mathcal{P}(t)< \! 0\,.
\end{dcases}	\label{eq:vh}
\end{align}
A similar setup was also discussed in a different context in \rcite{Amin:2018kkg} where it was argued that such rapid phase-transitions of the Higgs field can produce gravitational wave signals.
Note that the Higgs vev $v_h(\phi)$ vanishes at the end of the reheating period and the Higgs field remains in the symmetric phase until the electroweak phase transition at temperature scale $\op(100)\gev$. 
Therefore, masses of the Higgs-doublet components $h_i$ in the unbroken phase (${\cal P}(t)\!>\!0$) are $m_{h_i}^2(t) \!=\! |\mu_h^2(\phi)|$, whereas in the broken phase (${\cal P}(t)\!<\!0$) the would-be Goldstone bosons ($i=1,2,3$) disappear as longitudinal components of $W^\pm$ and $Z$ bosons while the physical Higgs boson mass $m_{h_0}^2(t) \!=\! 2|\mu_h^2(\phi)|$.
In \eq{eq:be_sm} we have assumed $\Gamma_{\phi}\!\simeq\!\Gamma_{\phi\to hh}$ so that $\phi\to hh$ nearly saturates the total inflaton decay rate while branching fractions to other decay channels, including dark sector, are assumed to be negligible.
Averaging over one period of inflaton oscillation, one finds 
\beq
\begin{aligned} 
 \Gamma_{\phi \rightarrow hh} & = \frac{g_{h\phi}^2}{32 \pi} \frac{\mpl^2}{m_\phi} \sum_{i=0}^{3} \bigg\langle {\rm Re}\bigg[\sqrt{1-\frac{4 m_{h_i}^2\!(t)}{m_\phi^2}}\,\bigg]\bigg\rangle \,, 
 \\
 &\simeq\Gamma_{\phi\to hh}^{(0)}
\begin{dcases}
1, & \langle m_{h_0}\rangle\! \ll\! m_\phi,	 \\
\frac{3\, m_\phi^2}{16\pi\, g_{h\phi}\mpl \,\varphi(a)}, & \langle m_{h_0}\rangle\!\gtrsim\! m_\phi,
\end{dcases}	
 \end{aligned}  \label{eq:Gam_phi}	
\eeq
where $\Gamma_{\phi\to hh}^{(0)}\!\equiv\!g_{h\phi}^2 M_{\rm Pl}^2/(32\pi\, m_\phi)$ is the inflaton partial width in the limit $m_{h_i}\!(t)\to 0$.
Note that time-dependence of $\Gamma_{\phi \rightarrow hh}$ through $m_{h_i}\!(t)$ is averaged over one period; therefore the width becomes a slowly-varying function of time $\propto a^{3/2}$.
This is an important result since when $4 m_{h_i}^2(t)/m_\phi^2\ll 1$, $\Gamma_{\phi \rightarrow hh}$ is just a constant. 
The variation of $\Gamma_{\phi \rightarrow hh}$  is illustrated in the lower panel of \fig{fig:plotn1gh5} (solid red curve)\,\footnote{Both figures in this work are generated adopting numerical (exact) solution of the Boltzmann equations. However, our approximate analytic results agree with the corresponding numerical ones to good precision.}; the maximal suppression is observed at the very beginning of reheating, then this suppression is gradually reduced, implying an increase of the width $\propto a^{3/2}$ eventually reaching $\Gamma_{\phi\to hh}^{(0)}$.

Here is a comment in order regarding calculating the time-averaged inflaton width when kinematical suppression is present in \eq{eq:Gam_phi}. Let us consider the 0-component of the Higgs doublet ${\bm h}$, i.e., the physical Higgs boson $h_0$. Note that the Higgs mass squared $m_{h_0}^2(t)\!=\!2|g_{h\phi}M_{Pl} \phi(t)|$ is $\phi$-dependent, therefore it oscillates between zero and some maximum value for each inflaton oscillation. As a result, there is always an interval during each inflaton oscillation when $m_\phi\!>\! 2 m_{h_0}$ and the inflaton width is non-zero, whereas during the remaining time interval of each oscillation when $m_\phi\!<\! 2 m_{h_0}$ the width vanishes. However, as we mentioned above, we time-average the width over one inflaton oscillation period, hence the time-averaged inflaton width is non-zero as given in \eq{eq:Gam_phi} which agrees well with the exact numerical result shown in \fig{fig:plotn1gh5}.

During the reheating phase, one can analytically solve the Boltzmann equation~\eqref{eq:be_sm} for the radiation energy density as 
\begin{align}
\rho_{\rm SM}^{}& \!\simeq\! 3\mpl^2 H_e \Gamma_{\phi\to hh}^{(0)} 
\begin{dcases}
\Big(\frac{a_e}{a}\Big)^{3/2}, & \langle m_{h_0}\!\rangle\! \ll\! m_\phi,	 \\
\!\!\frac{3\, m_\phi^2}{16\pi g_{h\phi}\mpl \,\varphi_e}, & \langle m_{h_0}\!\rangle\!\gtrsim\! m_\phi, 
\end{dcases}	
\end{align}
where $H_e \!\equiv\! \Lambda^2/(2\sqrt3 \mpl)\, {\rm sinh}^{-1}\!\big( 2/\sqrt{3\alpha} \big)$ is the Hubble scale at the end of inflation for the $\alpha-$attractor T-model~\eqref{eq:inf_pot}.
Note that kinematic effect due to a non-zero Higgs mass is crucial for the reheating dynamics, as manifestly shown in \fig{fig:plotn1gh5} where the solid (dashed) curves correspond to the cases when the inflaton-induced Higgs mass is included (neglected), implying time-dependent (constant) inflaton decay rate. In particular, if $m_{h_i}$ was negligible, one would obtain $\rho_{{\rm SM}}\! \propto \! a^{-3/2}$ and $T\! \propto\!\ a^{-3/8}$, whereas including the effect of the non-zero Higgs mass one would get $\rho_{{\rm SM}} \! \propto\! \ a^0$ and $T \! \propto\! a^0$.  It is seen from Fig.~\ref{fig:plotn1gh5} that the maximum temperature can remain approximately constant over $N_\text{kin}\!=\!\log(a_\text{kin}/a_e)\!\simeq\! 2/3 \log(g_{h\phi}\mpl^4/\Lambda^4)$ e-folds, which could be potentially important for ultraviolet freeze-in dark matter scenarios as we discuss in the following section. 

Before concluding this section, we want to emphasize that in our framework inflaton decays perturbatively to the SM Higgs bosons through the Higgs portal interaction \eqref{eq:Lhphi} during the reheating phase. For our allowed range of the inflaton-Higgs coupling $g_{h\phi}$~\eqref{ghphi_limit} we note that non-perturbative effects can potentially become relevant and lead to tachyonic production of Higgs modes, i.e., the so-called preheating regime. However, the Higgs potential~\eqref{eq:H_pot} also includes a self-interaction term with relatively large quartic coupling~$\lambda_h\!\sim\!0.1$. For large Higgs field values, the Higgs quartic interaction generates a large field-dependent Higgs mass, leading to significant suppression of the tachyonic Higgs production due to its $\phi$-dependent mass. Therefore, the dominant mechanism of energy transfer from the inflaton field to the Higgs field remains the perturbative decay, see also~\cite{Lebedev:2021tas}. 

Finally, let us also point out that the graviton coupling to the energy-momentum tensors of inflaton and the SM constitutes an irreducible portal providing a minimal contribution to the reheating process, as already noted in \cite{Clery:2021bwz,Haque:2022kez}. In principle, the inflaton field unavoidably transfers some of its energy density to the SM sector through graviton-mediated interactions. Such processes can be especially relevant at the onset of reheating when $\rho_\phi \sim \Lambda^4$ is the largest. Thus, it is essential to compare the gravitational contribution to the reheating process with that of direct inflaton-Higgs interactions. It turns out that the cubic inflaton-Higgs interaction~\eqref{eq:Lhphi} dominates if the inflaton-Higgs coupling satisfy the following condition $g_{h \phi} \!\gtrsim\!\frac{\rho_\phi }{\mpl^4}$, which is always satisfied for the considered range of $g_{h \phi}$, \eq{ghphi_limit}.

\paragraph{Production of DM during reheating.}
\label{DM-effects}
Here we consider an Abelian vector DM $X_\mu$ with mass $m_X$, generated via an Abelian Higgs mechanism with a large expectation value of a dark Higgs field $\Phi$ so that the radial Higgs mode is heavy and is integrated out.
We assume the following specific form of the interaction Lagrangian,
\beq
\begin{aligned}
\cL\supset& - \frac{h^{\mu\nu}}{\mpl} \Big[T_{\mu\nu}^{\sm}+T_{\mu\nu}^{\dm} + T_{\mu \nu}^\phi\Big]	\\
& -\frac{\cxp\, m_X^2}{2\mpl} \phi X_{\mu} X^{\mu}-\frac{\cxh\, m_X^2}{2\mpl^2} X_\mu X^\mu |{\bm h}|^2,	
\end{aligned}	 \label{eq:Lint}
\eeq
where the first term describes irreducible interactions between DM, the SM, and the inflaton $\phi$ through the graviton $h_{\mu\nu}$ portal, with $T_{\mu\nu}^{\rm{DM}}$, $T_{\mu\nu}^{\rm{SM}}$, and $T_{\mu\nu}^{\phi}$ being the energy-momentum tensors for the vector DM, the SM, and the inflaton $\phi$, respectively\,\footnote{Recent studies \cite{Mambrini:2021zpp,Haque:2021mab,Clery:2021bwz,Haque:2022kez,Clery:2022wib} have shown that the gravitational production of DM from the inflaton background through the graviton portal can dominate over the corresponding production from the SM annihilation.}. 
The second (last) term contains leading interactions of the vector DM with the inflaton (SM), which appears through higher-dimensional operators and, therefore, must be suppressed by the UV scale assumed here to be $\mpl$. 
Perturbativity sets an upper-bound on the Wilson coefficients $\cxp$ and $\cxh$ as
\begin{align}
\cxp &\!\lesssim\! \frac{\mpl}{\phi} \bigg(\frac{\Lambda}{m_X}\bigg)^2\!, &\cxh &\!\lesssim\! \min\!\bigg\{\bigg(\frac{\mpl}{\mx}\bigg)^2, \bigg(\frac{\mpl}{T_{\rm max}}\bigg)^2 \bigg\}. \label{c_limit}
\end{align}  

Note that the tree-level exchange of the graviton would lead to an interaction between the SM and DM proportional to $1/\mpl^2$, which is of the same order as the DM--SM Higgs effective operator with $\cxh\sim 1$. On the other hand, the lowest-order direct interaction between DM and the inflaton arises at the dim-5 level, and thus one can naively expect that gravitational interactions can be neglected in this case. 
However, as we show in the following, the gravitational interaction between the DM and inflaton can be dominant for large inflaton and DM masses (energy density). 
In the following, we treat gravitational production as a kind of irreducible ``background'' DM production mechanism, occurring in all considered scenarios. Let us also note that even though the gravitational scattering of the inflaton and SM particles with the DM seems to have a similar nature, they describe very distinct processes. In case of the ${\rm{SM}} \, {\rm{SM}}\! \rightarrow\! h_{\mu \nu}\! \rightarrow \!XX$ process, the DM particles are produced from the quantum SM fields, which are assumed to be in thermal equilibrium. Therefore, the SM annihilation to the DM through the graviton portal is sensitive to the temperature during the reheating phase. Contrarily, the inflaton ``annihilation'' to the DM through the graviton exchange proceeds from the vacuum in the background of the classical oscillating inflaton field, which is not in thermal equilibrium\,\footnote{The gravitational production of DM or SM quanta from the inflaton ``annihilation'' can be best understood as the graviton interaction with the classical oscillating inflaton condensate/background and then the graviton transferring the energy gained from the inflaton background to the DM or SM quanta.}. This DM production mechanism is sensitive to the energy density of the inflaton field and is independent of the temperature of the thermal bath. It is most efficient at the start of reheating phase when the inflaton energy density is the largest.

Apart from the gravitational production, in this work, we also consider two other mechanisms that populate the DM sector, (i) direct decays of the inflaton or the SM Higgs field, and (ii) freeze-in from SM particles that annihilate into the DM particles.
To track the evolution of the dark sector, it is convenient to define a comoving number density of DM as $N_X \equiv n_X a^3$, such that the corresponding Boltzmann equation is~\cite{Chung:1998rq},
\begin{align}
\frac{{\rm d}N_X}{{\rm d}a} &= \frac{\rho_\phi(a_e) a_e^3}{m_\phi}\frac{\Gamma_{\phi\to XX}}{H}  + \frac{a^2}{H} \Big[\mathcal{S}_\phi + \mathcal{S}_{\rm{SM}} + \mathcal{D}_{h_0}\Big] \,,  \label{eq:Xeq}
\end{align}
where $\Gamma_{\phi\to XX}$ is the time-averaged width of the inflaton field to the DM particles, while 
\begin{align}
\mathcal{S}_{\rm{SM}}  &\simeq  \bar{n}_X^2\langle \sigma\lvert v \rvert \rangle_{XX \rightarrow \rm{ SM \, SM}}  ,	 \qquad
\mathcal{D}_{h_0} \simeq \bar{n}_{h_0} \langle  \Gamma_{h_0 \rightarrow XX} \rangle,	\\
\mathcal{S}_\phi & = \frac{1}{128 \pi} \left( \frac{\rho_\phi}{\mpl^2}\right)^2 \left[1- \frac{m_X^2}{m_\phi^2} + \frac{ 3m_X^4}{4 m_\phi^4} \right]\sqrt{1- \frac{m_X^2}{m_\phi^2}}	\,.	\label{eq:Sphi}
\end{align}
Above  $\langle\cdots\rangle$ stand for thermal and time averaged quantity and $\langle \sigma\lvert v \rvert \rangle_{XX \rightarrow \rm{ SM \, SM}}$ is the thermally averged cross-section for the process $XX \rightarrow \rm{ SM \, SM}$, while $\bar{n}_{X/h_0}$ denotes equilibrium number density for $X/h_0$. The collision term $\mathcal{S}_\phi$ accounts for the gain of DM species in the $\phi \phi \!\rightarrow h_{\mu \nu} \!\rightarrow\! XX$ process at the expense of the inflaton field.
The source term $\mathcal{S}_{X}$ contains contributions from the annihilation of SM particles via s-channel graviton $h_{\mu \nu}$ and Higgs boson exchange as well as the Higgs boson annihilation via the effective operator $\sim X_\mu X^\mu |{\bm h}|^2$. The $h_0$ decay term $\mathcal{D}_{h_0}$ also emerges from the same effective operator when the Higgs field acquires the $\phi-$dependent vev. 
Above $\Gamma_{\phi\to XX}$ and $\Gamma_{h_0 \rightarrow XX}$ are inflaton and Higgs boson decay widths to DM, respectively, given as
\begin{align} 
\Gamma_{\phi \rightarrow XX} &\simeq \frac{\lvert \mathcal{C}_X^\phi \rvert^2}{128 \pi} \frac{m_\phi^3}{\mpl^2},  \label{eq:phi_decay}\\
\langle \Gamma_{h_0 \rightarrow XX} \rangle  &\simeq \frac{\lvert \cxh \rvert^2}{128 \pi} \frac{\langle v_h^2(\phi) \,m_{h_0}^3\rangle}{\mpl^4}\frac{K_1 \big(\tfrac{m_{h_0}}{T} \big)}{K_2 \big(\tfrac{m_{h_0}}{T} \big)}, \label{eq:h0_decay}  
\end{align} 
where the decays are kinematically allowed for \mbox{$m_{\phi,h_0}\!>\!2m_X$} and for brevity we neglect effects of order $m_X^2/m_{\phi,h_0}^2$. 
Above $K_{1,2}(x)$ are the modified Bessel functions of the second kind.

In what follows, we discuss each DM production mechanism separately, starting from the universal one --- the gravitational production. Keeping only gravitational contributions, one can easily integrate~\eqref{eq:Xeq}, starting at the end of inflation, and find a prediction for the present DM abundance. As mentioned above, the gravitational production of DM from the inflaton field is not thermal. Therefore for fixed inflaton potential parameters, i.e., $n,\alpha,\Lambda$, it mainly depends on the duration of the reheating period, i.e., the value of $g_{h \phi}$ which defines the efficiency of the reheating period. On the other hand, for the gravitational annihilation of SM particles to the DM, two non-trivial features affect the DM production, (1) in the presence of time-dependent inflaton decay width, the SM bath temperature $T(a)\sim \rho_{\rm SM}^{1/4}(a)$ is modified, and (2) for our specific model the $\phi$-dependent Higgs vev generates effective $\phi$-dependent masses for SM fermions and gauge bosons during reheating phase. 
These SM particle masses during the reheating period are rescaled by $v_h(\phi)/v_{\tiny{\rm  EW}}$ compared to their zero-temperature SM masses~\cite{Ahmed:2022tfm}. 
In this work, we adopt the EW value for the Higgs quartic couplings, i.e., $\lambda_h \simeq 0.13$. 
It is worth noticing that $m_{h_0}$ is determined by inflaton--Higgs interaction and the solution of the inflaton EoM~\eqref{eq:inf_eom} and does not depend on $\lambda_h$. 
On the other hand, masses of other SM particles are sensitive both to the inflaton--Higgs interaction and the choice of $\lambda_h$. 
The cross-sections for the SM scalars, fermions, and vectors annihilating through the s-channel graviton exchange into DM vectors can be found in the companion paper~\cite{Ahmed:2022tfm}, see also \cite{Garny:2017kha,Tang:2017hvq} where the SM masses are neglected. 
It is widely known that thermal gravitational production of the DM from the SM annihilation is effective only at the highest temperatures during reheating~\cite{Garny:2015sjg,Tang:2017hvq,Garny:2017kha}. Moreover, it turns out that the effective mass of the SM particles can exceed the thermal bath temperature, which leads to the Boltzmann suppression of DM production. Consequently, one expects that the DM particles are mainly produced from the gravitational annihilation of SM massless vectors.
As the temperature drops, gravitational annihilation becomes inefficient, and the comoving DM number density $N_X$ approaches some constant value at the end of reheating $a\simeq a_{\rm rh}$. This, in turn, can be easily related to the present DM relic abundance
\begin{align}
\Omega_X = \frac{m_X n_X(a_0)}{\rho_c}=  \frac{m_X}{\rho_c} \frac{N_X(a_{\rm rh})}{a_{\rm rh}^3} \frac{s_0}{s(a_{\rm rh})}, \label{eq:OmegaX}
\end{align}
where $s(a_{\rm rh})$ denotes the entropy density at end of reheating, while $\rho_c\!=\!1.054 \times 10^{-5} h^2\,  \rm{GeV\, cm^{-3}}$ and $s_0\!=\!2970\, \rm{cm^{-3}}$ are the critical density and the present-day entropy density, respectively.

\begin{figure*}[t!]
\begin{center}
\includegraphics[width=0.33\linewidth]{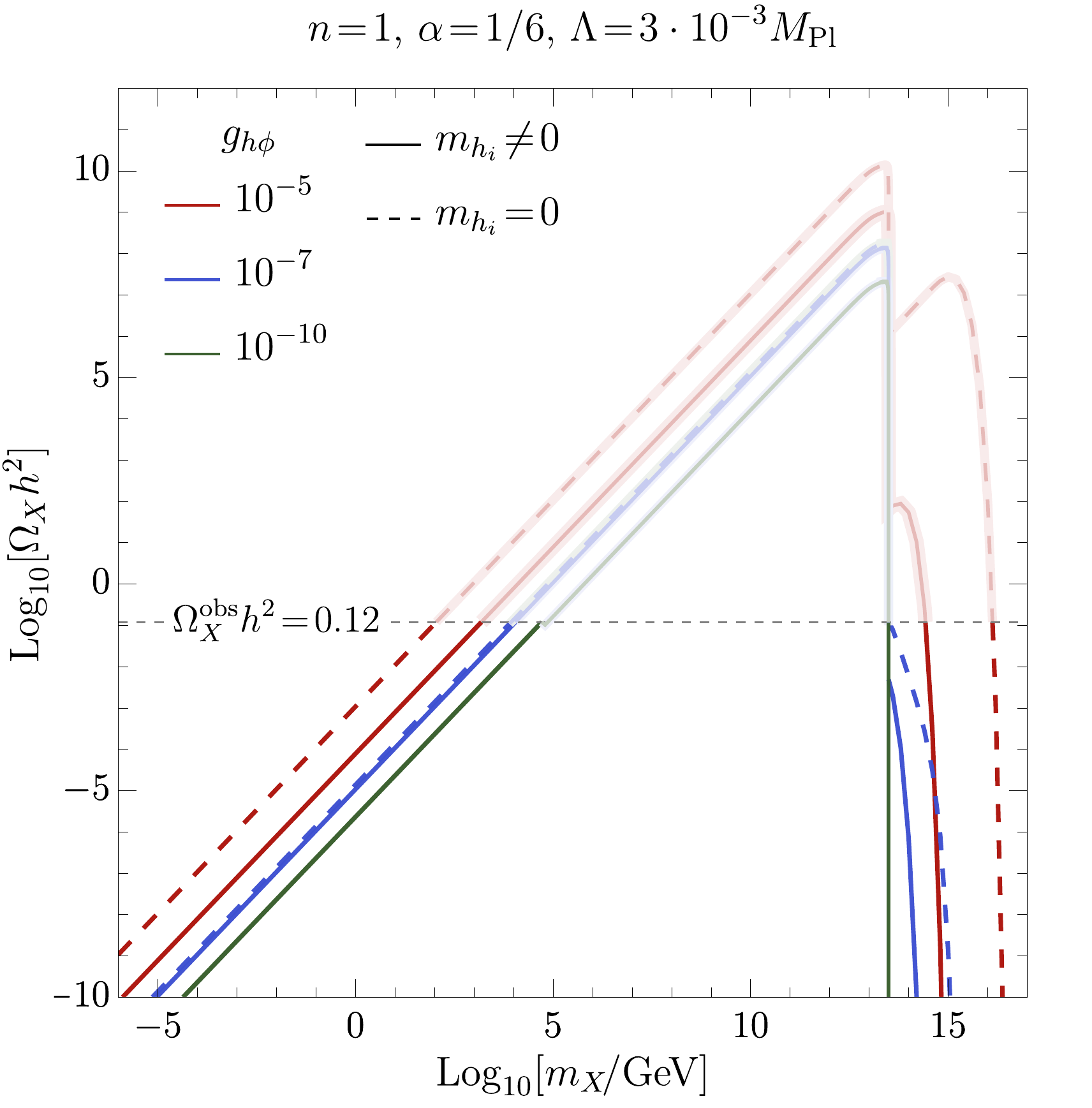} \!\!\! \includegraphics[width=0.33\linewidth]{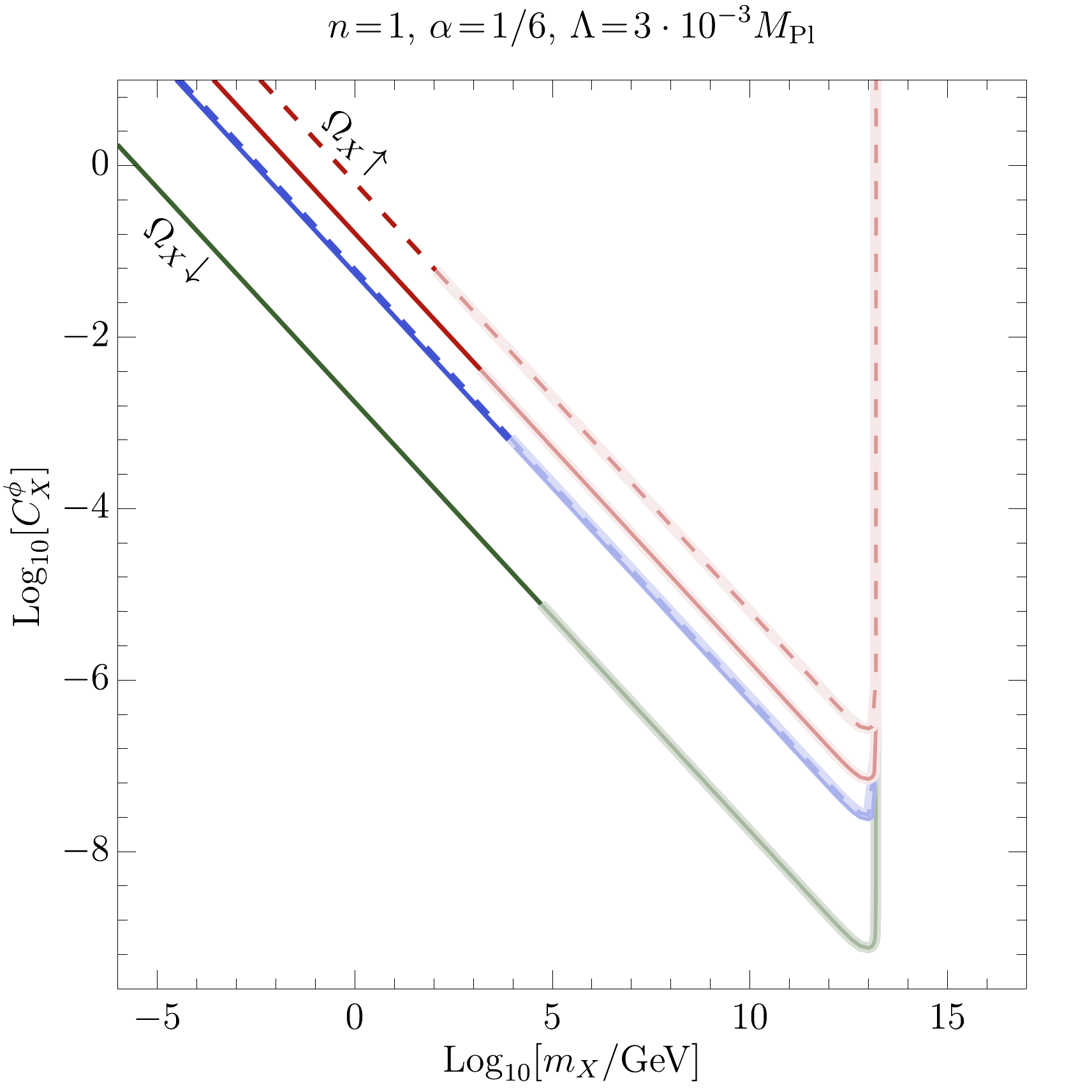}\!\!\!\includegraphics[width=0.33\linewidth]{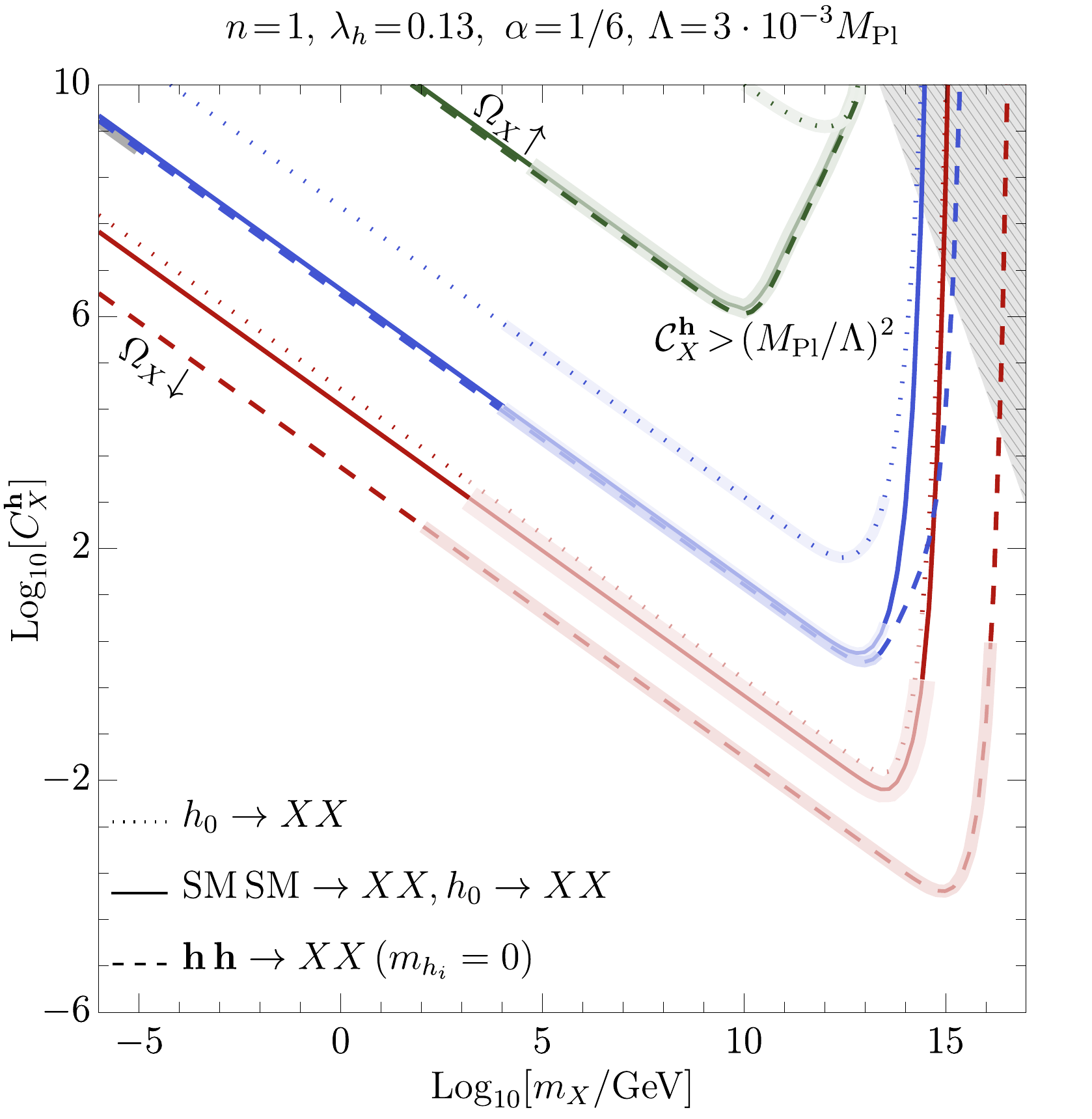}  \vspace{-3pt}
\caption{The left-panel shows DM production purely due to graviton exchange for massless (dashed, time-independent inflaton decay) and massive (solid, time-dependent inflaton decay) SM particles. The center and right panels respectively show the Wilson coefficients $\mathcal{C}_X^{\phi}$ and $\mathcal{C}_X^{\bm{h}}$ as a function of $\mx$ which give the observed DM abundance $\Omega_X^{\rm{obs}}h^2=0.12$. The dotted lines in the right-panel present the contribution from the Higgs boson decay to DM relic abundance. In hatched gray region $\mathcal{C}_X^{\bm{h}}$ exceeds the perturbativity limit $\cxh \gsim (\mpl/m_X)^2 $. The thick faded parts of these curves show values of $m_X$ where DM is overproduced due to gravitational productions only. The blue dashed curve in the right panel is also constrained by the perturbativity limit $\cxh \gsim (\mpl/T_{\rm max})^2 $ (the thick, gray part).}
	\label{fig:DMprod}
	\end{center}
\end{figure*}
In the left-panel of \fig{fig:DMprod} we show the present DM abundance $\Omega_X h^2$ due to only gravitational interactions as a function of the DM mass $m_X$ for different values of $g_{h\phi}$ which defines the reheating dynamics. The \textcolor{MyRed}{red}, \textcolor{MyBlue}{blue}, and \textcolor{MyGreen}{green} lines reproduce the observed DM relic for $g_{h\phi}$ equals $10^{-5}, 10^{-7}$ and  $10^{-10}$, respectively.  The solid (dashed) curves correspond to the cases when SM Higgs mass is included (neglected), implying time-dependent (constant) inflaton decay rate.
In \fig{fig:DMprod} we fix the inflationary parameters as $n\!=\!1$, $\alpha\!=\!1/6$, 
and $\Lambda\!=\!3\!\times\! 10^{-3}\mpl$ consistent with recent Planck data~\cite{Aghanim:2018eyx}, which results in the inflaton mass~\eqref{eq:mphi}, $m_\phi\simeq 3.1\!\times\! 10^{13}\gev$.  It turned out that there exists a wide range of DM mass shown in \fig{fig:DMprod} (left-panel) for which gravitational production of DM is relevant. We note that gravitational production from the inflaton ``annihilation'' is dominant over the corresponding SM annihilation for $m_\phi\!\geq\! m_X$~\eqref{eq:Sphi}. On the other hand, for larger DM mass, i.e. $m_X\!>\!m_\phi$ the inflaton gravitational production is shut down due to phase space, whereas the SM annihilation is allowed if the bath temperature is larger than the DM mass. This behavior is manifestly shown in the left-panel of \fig{fig:DMprod}. 
Let us also stress that the gravitational production of heavy DM species, i.e., $m_X\!>\!m_\phi$, from the annihilation of the SM particles, is significantly suppressed in the scenario with a massive Higgs field.
This could be understood by the fact that the maximal thermal bath temperature in the massive case is smaller than that of the constant inflaton decay rate case (for the massless Higgs scenario). 
Furthermore, due to non-trivial Higgs vev, the non-zero mass of the SM particles can exceed the temperature $T$, which leads to a Boltzmann suppression. Thus, in the massive reheating scenario, the DM production is dominated by the scattering of massless SM gauge bosons, while in the massless case, contributions from the annihilation of SM scalars, fermions, and vectors are approximate of the same order. Contrarily, the production of light DM, i.e., with mass  $m_X\!<\!m_\phi$, occurs mainly due to the gravitational  ``annihilation'' of the $\phi$ field and thus depends only on the inflaton potential parameters and the duration of reheating. The latter is affected by including (neglecting) the $\phi$-dependent Higgs mass effects, see the solid (dashed) curves. Note that pure gravitational vector DM production is excluded for a wide range of DM masses where $\Omega_X\!>\!\Omega^{\rm obs} \!=\! 0.12/h^2$~\cite{Aghanim:2018eyx}, see faded curves in \fig{fig:DMprod}.

Next, we discuss DM production through inflaton decays. The corresponding decay width averaged over inflaton oscillations is given in \eq{eq:phi_decay}.
In this case, DM particles are efficiently produced only during the reheating period when the $\phi$ field is still abundant. After reheating, the inflaton number density dilutes rapidly, and the $\phi \rightarrow XX$ channel is effectively turned off. The predicted relic abundance of DM species produced by the inflaton decay can be calculated from~\eqref{eq:OmegaX}. In \fig{fig:DMprod} (center-panel) we present the parameter space in the $\mathcal{C}_X^\phi-m_X$ plane that reproduces the observed DM relic abundance, i.e. $\Omega^{\rm obs} h^2 \!=\! 0.12$~\cite{Aghanim:2018eyx},for three benchmark values of the inflaton-Higgs coupling $g_{h \phi} =10^{-5}, 10^{-7}, 10^{-10}$. One can see that for smaller values of $g_{h \phi}$ one needs weaker inflaton-DM coupling $\mathcal{C}_{X}^{\phi}$ to satisfy $\Omega^{\rm{obs}}h^2$ constraint for a fixed DM mass. This behavior is related to the fact that for weaker inflaton-Higgs coupling the reheating lasts longer. Furthermore, the faded curves represent the parameter space where the DM is overproduced from the inflaton annihilations through the graviton portal. 

Next, we consider DM production during reheating due to decays of the massive Higgs boson. 
The DM-Higgs interaction term~\eqref{eq:Lint} when expanded around Higgs vev leads to the \mbox{$\mathcal{C}_X^{\bm{h}} m_X^2 v_h(\phi) h_0 X_{\mu}X^{\mu}/\mpl^2$} term, that accounts for the $h_0$ decay to DM vectors when $m_{h_0}\geq 2m_X$. The resulting time-averaged decay width is given in \eq{eq:h0_decay}.
It is worth noting that the $h_0 \rightarrow XX$ decay is possible only in one half of the inflaton oscillation period, i.e., when $v_h(\phi) \neq 0$, and it is switched off when reheating ends. 
The relation between the Wilson coefficient $\cxh$ and $m_X$ that predicts the correct DM abundance is presented in \fig{fig:DMprod} (right-panel, dotted curves), assuming that DM vectors are produced only through $h_0$ decays. Since $v_h(\phi)$, and thus $\Gamma_{h_0 \rightarrow XX}$, is proportional to $g_{h \phi}$, it is clear that for weaker inflaton-Higgs coupling, a larger value of $\cxh$ is required to produce the same amount of DM abundance with a fixed mass. 

Finally, we can discuss DM production from annihilations of SM particles either through the contact dim-6 operator, i.e. \mbox{$\cxh m_X^2 X_\mu X^\mu |{\bm h}|^2/\mpl^2 $} or through a single s-channel Higgs exchange.  Note that the former channel occurs in both massive and massless reheating scenarios, while the latter is present only in the massive one since a non-zero Higgs vev is needed.
The corresponding cross-sections for the SM scalars, fermions, and vectors annihilating into pairs of $X$ particles can be found in~\cite{Ahmed:2022tfm}. It turns out that for our choice of the $\lambda_h$ parameter, the source term that accounts for the $f \bar{f}$ annihilations into DM pairs is always smaller than $\mathcal{S}_{XX \rightarrow VV}$ or $\mathcal{S}_{XX \rightarrow h_i h_i}$. The final scan of $\mathcal{C}_{X}^{{\bm h}}$ and $m_X$ values consistent with observed DM relic is presented in \fig{fig:DMprod} (right panel, solid curves), where we have included all interactions that describe DM production through annihilations of SM particles either via contact operator or through s-channel Higgs exchange, as well as via Higgs decays. 
The dashed lines correspond to the case where Higgs mass is neglected (which implies $\Gamma_\phi$ constant), in this case, annihilations of the Higgs bosons occur via the contact operator.
The contribution of the $h_0$ decay to the total DM relic abundance is the largest for large inflaton-Higgs coupling $g_{h\phi}^{}$. As $g_{h \phi}^{}$ decreases, the $h_0 \rightarrow XX$ channel becomes less and less relevant for DM production. For all considered values of $g_{h \phi}^{}$, dashed curves lie below corresponding solid ones, which means that in the massless reheating scenario, one needs a smaller value of the Wilson coefficient $\cxh$ to produce DM particles with a given mass in the correct amount. The discrepancy between the massive and massless case is most significant for larger values of $g_{h \phi}$. 
This is because for large inflaton-Higgs coupling the thermal bath temperature $T(a)$ is suppressed due to kinematic suppression during the reheating phase and hence causing kinematical suppression of DM production. Again the faded thick curves of the parameter space are excluded due to gravitational overproduction of DM discussed above, whereas the hatched region, $\cxh\!>\!\mpl^2/\Lambda^2$, is not allowed due to perturbativity~\eqref{c_limit}. 

\paragraph{Conclusions.}
We have discussed the implications of time-dependent inflaton decay for 
radiation and dark matter production during the reheating phase. We presented general results for the SM radiation production when the inflaton decay rate has power-law dependence on the scale factor. Such a time-dependent inflaton decay rate can be naturally realized in models where inflaton interacts with the SM Higgs boson. As an example, we considered the interaction of the form $\propto \phi |{\bm h}|^2$ between the inflaton $\phi$ and SM Higgs doublet  ${\bm h}$. This interaction generates the $\phi$-dependent Higgs boson mass/vev, leading to a kinematical suppression of inflaton decay to the Higgs boson; hence the inflaton decay rate becomes time-dependent. This phenomenon modifies the dynamics of the reheating period, affecting the evolution of the radiation energy density and thus the temperature of the thermal bath. It has also been pointed out that the DM freeze-in production strongly depends on the details of the reheating. In particular, it has been demonstrated that the non-zero mass of the Higgs boson not only has a significant impact on gravitational DM production but it also provides new DM production channels. In each scenario, the viable region in parameter space consistent with the observed amount of DM has been found.
\paragraph{Acknowledgments.}
\noindent The work of B.G. and A.S. is supported in part by the National Science Centre (Poland) as a research project, decisions no 2017/25/B/ST2/00191 and 2020/37/B/ST2/02746.

\bibliographystyle{aabib} 
\bibliography{bib_vdm_reh}{} 
\end{document}